\documentstyle[preprint,eqsecnum,aps,epsf]{revtex}	

\newif\iftightenlines\tightenlinesfalse
\tightenlines\tightenlinestrue

\begin{document}
%
\def\mol{Mol}
\def\eslt{E\llap/_T}
\def\esl{E\llap/}
\def\msl{m\llap/}
\def\to{\rightarrow}
\def\te{\tilde e}
\def\tmu{\tilde\mu}
\def\ttau{\tilde\tau}
\def\tl{\tilde\ell}
\def\ttau{\tilde \tau}
\def\tg{\tilde g}
\def\tnu{\tilde\nu}
\def\tell{\tilde\ell}
\def\tq{\tilde q}
\def\tu{\tilde u}
\def\tc{\tilde c}
\def\tb{\tilde b}
\def\tst{\tilde t}
\def\tt{\tilde t}
\def\tw{\widetilde W}
\def\tz{\widetilde Z}

\hyphenation{mssm}
%
\preprint{\vbox{\baselineskip=14pt%
   \rightline{FSU-HEP-961001}\break 
   \rightline{hep-ph/9610224}\break 
}}
\title{QCD IMPROVED $b\to s\gamma$ CONSTRAINTS ON THE\\
MINIMAL SUPERGRAVITY MODEL}
\author{Howard Baer and Michal Brhlik}
\address{
Department of Physics,
Florida State University,
Tallahassee, FL 32306 USA
}
\date{\today}
\maketitle
\begin{abstract}

Recent advances in the QCD corrections to $b\to s\gamma$
decay in the MSSM include {\it i.}) evaluation of the relevant operators,
Wilson coefficients and anomalous dimension matrix elements for the various 
MSSM effective theories valid at scales beyond $Q =M_W$, {\it ii.})
calculations of most of the needed anomalous dimension matrix elements
to next-to-leading order for scales $m_b\alt Q <M_W$, and
{\it iii.}) calculations of ${\cal O}(\alpha_s)$ virtual and bremsstrahlung
corrections to the $b\to s\gamma$ decay operators at scale $Q\sim m_b$.
We assemble all these known results to gain an estimate of 
$B(b\to s\gamma )$ for the 
parameter space of the minimal supergravity model (mSUGRA). We find a 
much reduced scale dependence of our result compared to usual 
leading-log evaluations.
Comparison with the latest CLEO results yields stringent constraints
on parameter space. Much of mSUGRA parameter space is ruled out for $\mu <0$,
especially for large $\tan\beta$. We compare these results with 
other constraints from cosmology and non-standard vacua. Also, we 
compare with expectations for discovering mSUGRA at LEP2, the
Tevatron and the CERN LHC.

\end{abstract}

\medskip



\section{Introduction}

Particle physics models including weak-scale supersymmetry (SUSY) are 
amongst the most compelling candidates\cite{haber} for physics beyond 
the Standard Model (SM). Of this class of models, the minimal 
supergravity model (mSUGRA) stands out as at least the most popular 
framework for performing searches for SUSY, and can justifiably be 
called the paradigm model for weak scale supersymmetry\cite{drees}. 
The mSUGRA model 
can be characterized briefly by the following attributes\cite{sugra,drees}:
\begin{itemize}
\item particle content and gauge symmetries of the 
Minimal Supersymmetric Standard Model (MSSM), {\it i.e.} a supersymmetrized
version of the two-Higgs doublet SM, plus all allowed soft SUSY breaking terms,
\item a desert between the weak scale and the unification scale, which allows
for gauge coupling unification,
\item universal boundary conditions for soft-SUSY breaking terms
$m_0,\ m_{1/2},\ A_0$ and $B_0$ implemented at the unification scale $M_X$,
\item electroweak symmetry is broken radiatively, and is a consequence 
of the large top quark Yukawa coupling.
\end{itemize}
The various weak scale parameters are related to GUT scale parameters via 
renormalization group equations (RGE's). Typically, all weak-scale
sparticle masses and mixings are then determined by the parameter set
\begin{equation}
m_0,\ m_{1/2},\ A_0,\ \tan\beta\ {\rm and}\ sign(\mu ),
\label{eq1}
\end{equation}
along with the measured value of $m_t$. We take $m_t=170$ GeV throughout
this paper.

Of course, not all values of the above parameter set are allowed. For some
values, electroweak symmetry is not broken appropriately. For other values,
a charged or colored sparticle may be the lightest SUSY particle (LSP), in
conflict with cosmology and searches for exotic nuclei and atoms. In addition,
there exist constraints from negative searches for sparticles at the 
Fermilab Tevatron $p\bar p$ and LEP2 $e^+e^-$ colliders. 
In particular, we note the 
recent bound that $m_{\tw_1}>79$ GeV for a gaugino-like lightest 
chargino\cite{lep2}.
Additionally, in the absence of $R$-violating interactions, there exist
bounds on parameter space from the relic density of neutralinos 
produced in the Big Bang\cite{cosmo}; 
for certain regions of parameter space, the 
neutralino relic density $\Omega h^2>1$, which implies a universe younger than
10 billion years, in contradiction at least with the ages of the oldest 
stars in globular 
clusters. Finally, several recent papers have mapped out regions of mSUGRA
parameter space where there exist vacua deeper than the standard 
minimum\cite{vacua}. These non-standard vacua 
constraints may not be rigorous, however, if one 
entertains the notion that the universe may have settled into a false vacuum.

The parameter space of the mSUGRA model (as well as many other 
models\cite{hewett}) 
may also be constrained by data from rare meson decays, such as 
the branching fraction for $B\to X_s\gamma$. 
A recent analysis by the CLEO collaboration finds
for the inclusive decays $B(B\to X_s\gamma )=(2.32\pm 0.67)\times 10^{-4}$,
with 95\% CL upper and lower limits (including systematic errors) 
of $4.2\times 10^{-4}$ and $1\times 10^{-4}$, respectively\cite{cleo}. 
Such data, when compared against theoretical predictions
of $b\to s\gamma$, have been shown to be very 
restrictive for both two-Higgs doublet models (2HDM)\cite{higdoub} 
and supersymmetric models\cite{masiero,susybsg}.
For a type II 2HDM, loop contributions involving the top quark and 
charged Higgs $H^\pm$
add constructively with SM loops involving top and $W$ bosons. In the MSSM,
there exist other contributions involving squark-chargino loops,
squark-neutralino loops, and squark-gluino loops. The latter two
contributions are much smaller than squark-chargino loop contributions, and 
are frequently neglected. The squark-chargino loop contribution can add
either constructively or destructively with the $W$ and $H^+$ loops, leading 
to allowed or forbidden regions of SUSY model parameter space.

These calculations are usually performed by evaluating lowest order 
matrix elements of effective theory operators at a scale $Q \sim m_b$. 
All orders approximate QCD corrections are included via renormalization group
resummation of leading logs (LL) which arise
due to a disparity 
between the scale at which new physics enters the $b\to s\gamma$ loop
corrections (usually taken to be $Q\sim M_W$), and the scale at which 
the $b\to s\gamma$ decay rate is evaluated ($Q\sim m_b$). 
The resummation is most easily performed within the framework of
effective field theories. Above the scale $Q =M_W$ (all scales $Q\sim M_W$
are equivalent in LL perturbation theory), calculations
are performed within the full theory. Below $Q =M_W$, heavy particles are
integrated out of the theory,
leading to an effective Hamiltonian
\begin{equation}
H_{eff}=-{4G_F\over \sqrt{2}} V_{tb}V^*_{ts}\sum_{i=1}^8 C_i(Q )O_i(Q ),
\label{eq2}
\end{equation}
where matching between the two theories occurs at $Q =M_W$ and yields the
values of $C_i(Q =M_W)$. In 
Eq. \ref{eq2}, the $C_i(Q )$ are Wilson coefficients evaluated at scale $Q$,
and the $O_i$ are a complete set of operators, given, for example, 
in Ref. \cite{gh}. Resummation then occurs by solving the renormalization
group equations (RGE's) for the Wilson coefficients
\begin{equation}
Q {d\over dQ} C_i(Q )=\gamma_{ji} C_j(Q ),
\label{eq3}
\end{equation}
where $\gamma$ is the $8\times 8$ anomalous dimension matrix (ADM), 
and
\begin{equation}
\gamma={\alpha_s\over 4\pi}\gamma^{(0)}+({\alpha_s\over 4\pi})^2 
\gamma^{(1)}+\ldots .
\label{eq4}
\end{equation}
The matrix elements of the operators
$O_i$ are finally calculated at a scale $Q\sim m_b$ and multiplied by
the appropriately evolved Wilson coefficients to gain the final 
decay amplitude.

The LL QCD corrections just described yield enhancements in the 
$b\to s\gamma$ decay rate of factors of
$2-5$\cite{llqcd,pokorski}. The resulting LL calculation yields an answer which 
is ambiguous depending upon which precise scale choice is chosen for
evaluation of matrix elements of the operators $O_i$.
Variation of the scale ${m_b\over 2}< Q < 2m_b$ yields approximately a
25\% uncertainty in the theoretical calculation. This uncertainty 
provides the greatest source of error in currently available
theoretical calculations\cite{pokorski}. To reduce the theoretical
uncertainty, one must proceed to a next-to-leading-log calculation (NLL)
of the $b\to s\gamma$ decay rate.

Recently, a number of theoretical developments have been made towards the 
goal of a NLL calculation of $B(b\to s\gamma )$. Cho and Grinstein
showed\cite{cg} that if there are two significantly different masses 
contributing to the loop amplitude
(such as $m_t$ and $M_W$), then in fact there can already exist significant
corrections to the Wilson coefficients at scale $M_W$. In this case, 
one must create an effective theory by first integrating out the 
heavy top quark, apply RGE running between the scales $m_t$ and $M_W$, and then
integrate out the $W$ boson to arrive at the operator set in \ref{eq2}.
The above procedure gives a $\sim 20\%$ enhancement to the SM value of
$B(b\to s\gamma )$. In supersymmetric models, many more heavy particles are 
present. Anlauf has shown how to perform a similar analysis for the case 
of the MSSM\cite{anlauf}. These corrections are considered to be 
next-to-leading order effects.

In addition, various terms of the ADM in Eq. \ref{eq4} have been calculated 
to next-to-leading order (NLO). The ${\cal O}(\alpha_s^2)$ terms $\gamma_{ij}^{(1)}$
for $i,j=1-6$ have been calculated by 
Ciuchini {\it et. al.}\cite{ciuchini}, 
while the corresponding terms for $i,j=7,8$ are given by 
Misiak and M\"unz\cite{misiak1}. Of the remaining terms which mix the
$i=1-6$ and $j=7,8$ Wilson coefficients, only $\gamma_{27}^{(1)}$ is
relevant, but its evaluation requires a three loop calculation.
A preliminary report on the calculation of $\gamma_{27}^{(1)}$ 
indicates that it is only a small effect\cite{misiak2}.

Finally, the QCD corrections to the operators $O_i$ must be 
included\cite{soares,ghw}.
Recently, Greub, Hurth and Wyler have reported results\cite{ghw} on the
complete virtual corrections to the relevant operators $O_2,\ O_7$ and $O_8$.
Combining these with the bremsstrahlung corrections\cite{brem,ghw} 
results in cancellation of associated soft and collinear singularities.
A combination of the QCD corrected operator matrix elements $<s\gamma |O_i|b>$
with the complete ${\cal O}(\alpha_s)$ corrected Wilson coefficients 
at scale $Q\sim m_b$ will result in a NLL calculation of the 
$B(b\to s\gamma )$ decay rate.

In this paper, we have nothing new to add to the calculational procedure
for evaluating $B(b\to s\gamma )$.
Our goal in this paper is to bring together the above pieces of a 
NLL calculation of $B(b\to s\gamma )$ and to interface with the mSUGRA
model so that detailed comparisons to data can be made in parameter space.
In so doing, we simply neglect the missing piece of the calculation,
$\gamma_{27}^{(1)}$. This will result in some small scheme dependence of 
our results, and in some additional scale dependence, so our calculation will
not be truly NLL. Hence, we label it as QCD improved, in hope that 
$\gamma_{27}^{(1)}$ turns out small, as preliminary reports suggest.

Our goal as well is to evaluate the $B(b\to s\gamma )$ rate as a function
of mSUGRA parameters and compare with the recent CLEO results, to find
favored or excluded regions of parameter space. We find the 
$b\to s\gamma $ constraint to be really very strong, as noted 
previously\cite{masiero,susybsg}. 
We compare the $b\to s\gamma$
results to other recent results on relic density constraints\cite{cosmo}
and regions of non-standard vacua\cite{vacua}. Finally, we note the effect
of $B(b\to s\gamma )$ on expectations for discovering mSUGRA at various 
collider experiments. Toward these ends, in Sec. 2 we present various 
details of our QCD improved calculation for $B(b\to s\gamma )$. In Sec. 3,
we report on our results as a function of mSUGRA parameter space, and compare
with other constraints and expectations for collider searches. We summarize
in Sec. 4.

\section{Calculational method}

In this section, we outline our procedure for calculating 
$B(b\to s\gamma )$ as a function of mSUGRA parameter space. Our first step
of course is to input the parameter set \ref{eq1} and solve for
the superparticle masses and mixings via running of the mSUGRA
RGE equations between $M_Z$ and $M_{GUT}$ and imposing the appropriate
minimization criteria using the 1-loop corrected effective potential.
We iterate the running back and forth between the two scales six times
using the usual Runge-Kutta method; this results in a convergent 
spectrum of superparticle masses.
The procedure is described more fully in Ref. \cite{bcmpt}.  
On the last run down but one, we take note of the various 
superparticle masses. On the final run from $M_{GUT}$ to $M_Z$, 
we implement the procedures outlined in Anlauf\cite{anlauf} to 
ultimately obtain the needed Wilson coefficients $C_7(M_W)$ and $C_8(M_W)$.
In addition, the values of $C_i(M_W)$ for $i=1-6$ are given in 
Ref. \cite{ciuchini} to ${\cal O}(\alpha_s )$.

Given a heavy particle of mass $m_H$ and a light particle of mass $m_L$
contributing to a $b\to s\gamma$ loop, the procedure of Anlauf is as follows.
First, at scale $Q =m_H$, the heavy particle is decoupled from the theory
and the corresponding effective field theory is constructed. The leading terms
of the expansion of $C_i$ in terms of $x=({m_L\over m_H})^2$ are calculated 
and evolved to scale $Q =m_L$ using the ADM including both 
QCD and electroweak interactions. 
Taking only leading terms in $x$ restricts the operator basis to dimension
5 and 6 operators\cite{anlauf}.
At $Q =m_L$, the remaining part of $C_i$,
which has been evolved down to $m_L$ using only the EW ADM, is added together
with eventual contributions coming from decoupling of the lighter particle 
in the loop. As a last step, the equations of motion are applied to obtain
$C_7(M_W)$ and $C_8(M_W)$. In the case that $m_L<M_W$, the evolution is 
stopped at $M_W$. In practise, this is almost never a problem, in light 
of the new bound $m_{\tw_1}>79$ GeV for gaugino-like charginos from 
LEP2. In our calculation, we include contributions from $tW$, $tH^-$
and $\tw_i \tq_j$ loops, for $i=1$ and 2, and 
$\tq_j=\tu_L,\ \tu_R,\ \tc_L,\ \tc_R,\ \tst_1$ and $\tst_2$, and neglect
contributions from $\tz_i\tq_j$ and $\tg\tq_j$ loops, which should have
much smaller contributions. 

As an illustration of these results, we show in Fig. 1 the final value of
the Wilson coefficient $C_7(M_W)$ from the various loop 
contributions just discussed, as a function of $m_{1/2}$, where
$m_0=m_{1/2}$, $\tan\beta =2$ and $A_0=0$. In frame {\it a}) $\mu <0$,
while in frame {\it b}), $\mu >0$. The $tW$ SM 
contribution is just $\simeq -0.23$, and of course doesn't vary versus
SUSY soft-breaking parameters. In both frames, the $tH^-$ contribution 
is negative, and 
decreases in absolute value as $m_{1/2}$ increases, since the value
of $m_{H^-}$ is increasing. Note this contribution adds constructively to the
SM $tW$ contribution; when combined, these give the large constraints 
on type II 2HDM's\cite{higdoub}.
For $\mu <0$, most of the SUSY particle loop contributions are also
negative in this case, which leads to the significant constraints 
to be given in Sec. 3. The exceptions are the large positive contributions 
from $\tw_2\tst_2$ and $\tw_1\tq$, which cancel 
some of the large negative contributions.
Alternatively, for $\mu >0$, we see in frame {\it b}) that there are several
positive as well as negative contributions to $C_7(M_W)$. In this case,
one can achieve rates for $B(b\to s\gamma )$ which are equal to or even smaller
than the SM value.

The next step in our calculation is to implement the NLO ADM to 
calculate the running of the $C_i$'s from $M_W$ down to $Q\sim m_b$.
The terms $\gamma_{ij}^{(1)}$ for $i,j=1-6$ have been calculated 
in Ref. \cite{ciuchini}. The corresponding ADM elements for $i,j=7,8$
have been calculated in Ref. \cite{misiak1}. The remaining NLO 
ADM elements for $\gamma_{ij}^{(1)}$ for $i=1-6$ and $j=7,8$ have not yet 
been published. The most important of these is $\gamma_{27}^{(1)}$, 
since $C_2(M_W)\simeq 1$ while the remaining $C_i(M_W)$ which mix into
$C_7$ are $\sim 0$\cite{ghw}. Preliminary results of the three-loop
calculation of $\gamma_{27}^{(1)}$ indicate it is small\cite{misiak2},
so in our calculation, we take $\gamma_{27}^{(1)}=0$. 

We show in Fig. 2 the evolution of the set of Wilson coefficients
$C_i$ for $i=1-8$, from their values at $Q =M_W$ down to $Q =1$ GeV.
Frame {\it a}) shows the evolution including just LO terms in the ADM, while
frame {\it b}) includes the NLO ADM contributions mentioned above.
We see that in general, the NLO effects are small. The exception is for 
$C_7$, for which the NLO correction differs from LO by $\sim 10\%$. 

Finally, we must include the matrix elements $<s\gamma |O_i(Q)|b>$
at NLO. 
We neglect the $O_3$, $O_4$, $O_5$ and $O_6$
contributions, since the corresponding $C_i$ are small, as shown in Fig. 2.
Furthermore, the matrix element $<s\gamma |O_1|b>$ is exactly zero.
The bremsstrahlung graphs have been calculated in 
Ref. \cite{brem} for $b\to s\gamma g$, while the complete 
virtual corrections to the 
operators $O_2,\ O_7$ and $O_8$ have been 
calculated in Ref. \cite{ghw}. 
We implement
these results, which ensure the proper cancellation of infrared and 
collinear singularities. Since our final results are not completely NLL, 
we will have some remaining scheme dependence. All our calculations have
been performed within the naive dimensional regularization (NDR) scheme,
in which the calculational building blocks have been given.
We neglect throughout our calculation any long distance effects\cite{desh} 
on the $b\to s\gamma$ decay rate.

Our final numerical result is given by
\begin{equation}
B(b\to s\gamma )={\Gamma\over \Gamma_{sl}} B_{sl},
\label{2.1}
\end{equation}
where
\begin{equation}
\Gamma (b\to s\gamma )=\Gamma_{virt} +\Gamma_{brem}.
\label{2.2}
\end{equation}
In the above, $\Gamma_{virt}$ is given by Eq. (5.6) of Ref. \cite{ghw}, and
$\Gamma_{brem}$ is given in Ref. \cite{brem,ghw}, while $\Gamma_{sl}$ is given
by Eq. (5.9) of Ref. \cite{ghw}. Numerically, we take the combination
${{V_{ts}^* V_{tb}}\over {|V_{cb}|^2}}=0.95$ and $B_{sl}=0.104$.

The inclusion of the various above mentioned QCD improvements leads to a 
result for $B(b\to s\gamma )$ which has significantly reduced 
scale-dependence.
We illustrate the scale dependence of our result in Fig. 3, where we plot
$B(b\to s\gamma )$ versus $Q$, where ${m_b\over 2}<Q <2m_b$. The 
LL calculation for the SM value is denoted by the dashed curve, which
yields $B(b\to s\gamma )=2.9\pm 0.7$, or a 25\% uncertainty due to 
scale choice. The QCD-improved result is shown by the solid curve.
In this case, the prediction is $B(b\to s\gamma )=3.2\pm 0.3$, and the error
due to scale choice uncertainty is reduced to $\sim 9\%$. The CLEO measured
central value is denoted by the solid horizontal line. The $1\sigma$ limits
are denoted by dotted lines, and the 95\% CL limits are denoted by
solid lines. We see that the SM prediction lies somewhat above the CLEO
measured result, although it is well within the 95\% CL region.

\section{Numerical results and Implications}

\subsection{Results for $B(b\to s\gamma )$}

We present our main results on the $b\to s\gamma$ branching ratio as
contours of constant branching fraction in the $m_0\ vs.\ m_{1/2}$
plane. This allows a direct comparison of previous work on 
mSUGRA constraints and also collider expectations to be made with the 
present work. In Fig. 4, we show the $B(b\to s\gamma )$ contours 
for $A_0=0$ for {\it a}) $\tan\beta =2,\ \mu <0$, 
{\it b}) $\tan\beta =2,\ \mu >0$, {\it c}) $\tan\beta =10,\ \mu <0$ and
{\it d}) $\tan\beta =10,\ \mu >0$. Each contour must be multiplied by $10^{-4}$.
The values of $B(b\to s\gamma )$ shown are for $Q =m_b$.
The regions marked TH are excluded by theoretical constraints: 
either the lightest SUSY particle (LSP) is not the neutralino $\tz_1$, 
or electroweak symmetry is improperly broken. The regions marked EX are
excluded by negative SUSY particle search experiments at Fermilab Tevatron 
or LEP. The new bound on $m_{\tw_1}>79$ GeV on gaugino-like charginos 
from LEP2 is indicated by the dashed contour.

In frame {\it a}), we find $B(b\to s\gamma )$ to be large for 
small values of $m_0$ and $m_{1/2}$. This is due to constructive 
interference amongst many of the SUSY and SM loop contributions, as can be
gleaned from Fig. 1{\it a}. As $m_0$ and $m_{1/2}$ increase, 
the SUSY particles and charged Higgs boson all increase in mass, and their
loop contributions become small: thus the value of $B(b\to s\gamma )$
gradually approaches its SM value in the upper-right region of each frame.
We note the CLEO 95\% CL upper bound on the inclusive rate for
$B(b\to s\gamma )$ is $4.2\times 10^{-4}$, so it becomes evident that 
a significant region to the lower-left of frame {\it a}) will become 
excluded. In frame {\it b}), for $\mu >0$, the results are significantly 
different. In this case, there are many interfering loop contributions
(see Fig. 1{\it b}),
so the $B(b\to s\gamma )$ rate is much closer to the 
SM value, and can even drop below it.
Since all the contours lie within the CLEO 95\% excluded band, this
frame remains unconstrained by $B(b\to s\gamma )$. For frame {\it c}),
with $\tan\beta =10$ and $\mu <0$, we find very large values of 
$B(b\to s\gamma )$ throughout the entire region of the plane shown. 
Almost all of the plane
shown gives values of $B(b\to s\gamma )$ greater than the CLEO 95\% CL
bound, and so will be excluded. Values of $m_{1/2}\agt 500$ GeV 
are required to reach an allowed region for this choice
of mSUGRA parameters. Finally, in frame {\it d}), we show the
$B(b\to s\gamma )$ result for $\tan\beta =10$ and $\mu >0$. As in 
frame {\it b}), there exists substantial interference amongst the various loop
contributions. The interference in this case is so great that a large 
fraction of the plane actually has $B(b\to s\gamma )$ values 
{\it below} the SM value. 
The $B(b\to s\gamma )$ rate increases with $m_{1/2}$ to reach its SM value.

Fig. 4 showed results for the GUT scale trilinear coupling $A_0=0$. 
Changing the value of $A_0$ will change the weak scale $A$ parameters,
which can result in changes to the top squark mass matrix and mixing angles.
This can then affect the $\tw_i\tst_j$ loop contributions. To show the effect
of changing $A_0$, we show in Fig. 5 the $B(b\to s\gamma )$ contours in
the $m_0\ vs.\ A_0$ plane, for fixed $m_{1/2}=200$ GeV, and all other 
parameters as in Fig. 4. There are some small TH excluded regions in the 
left-hand corners of the $\tan\beta =2$ frames. In frame {\it a}), we see
that the $B(b\to s\gamma )$ rate varies mainly with $m_0$ rather than
$A_0$, and that only a small portion of the plane is above the CLEO
95\% excluded value of $B(b\to s\gamma )=4.2\times 10^{-4}$.
In frame {\it b}), the $B(b\to s\gamma )$ rate varies slowly versus 
parameters, and the entire plane shown is allowed. Frame {\it c}) again 
varies slowly with $A_0$, and is entirely exluded. Frame {\it d}) 
has significant
variation against parameters, but is still entirely allowed.

\subsection{Relationship to Other Constraints on the mSUGRA Model}

Our next task is to compare the constraints from $b\to s\gamma$ with other
constraints, and derive conclusions relevant for collider searches.
Towards this end, we show in Fig. 6 regions of the $m_0\ vs.\ m_{1/2}$
plane which are excluded by CLEO data on $B(b\to s\gamma )$ at 95\% CL.
We match against the theoretical result from this paper. To obtain the 
excluded region, the $B(b\to s\gamma )$ rate must fall outside the CLEO
allowed values for {\it all} scale choices ${m_b\over 2}<Q <2m_b$.
The relevant excluded region for frame {\it a}) lies to the lower-left of 
the solid contour labelled $b\to s\gamma$. We show as well the dashed 
contour, below which $m_{\tw_1}<79$ GeV, in violation of recent 
LEP2 chargino searches\cite{lep2}. The region to the left of the line 
of open circles
is where non-standard minima of the mSUGRA model scalar potential 
lie\cite{vacua}.
This region may not be truly constrained if one is willing to accept 
that our universe may have settled into a false vacuum. 
The region to the right of the solid contour labelled by $\Omega h^2$
is excluded by cosmological considerations\cite{cosmo}. In $R$-conserving
models where the lightest neutralino is the LSP, it can be an excellent
candidate to make up the bulk of the dark matter of the universe. Such
a cold dark matter (CDM) particle would have been 
produced in abundance in the early universe,
and their present day abundance can then be straightforwardly 
calculated. If the relic density is too high, then the calculated 
lifetime of the 
universe is too short to be consistent with various astrophysical
observations. The contour denoted by $\Omega h^2$ denotes where
the relic density $\Omega h^2=1$; for higher values of $\Omega h^2$,
the universe would be younger than 10 billion years old, and so is 
certainly excluded. In addition, a number of cosmological observations
(COBE data, nuclear abundances, large scale structure) favor a universe
formed with a 2:1 ratio of CDM to hot dark matter (HDM- {\it e.g.} neutrinos).
This is called the mixed dark matter scenario (MDM).
This cosmologically favored region lies between the dot-dashed
contours, for which $0.15<\Omega h^2<0.4$. 

We combine the above mentioned constraints all on one plot, as in Fig. 6.
We see that in frame {\it a}), if one combines the false vacua region with
the $B(b\to s\gamma )$ constraint, much of the cosmologically favored 
MDM region is ruled out! 
Recent calculation of parameters associated with fine-tuning in the mSUGRA
model\cite{diego} actually prefer the small $m_0$ and $m_{1/2}$ regions that are
excluded by $B(b\to s\gamma )$ in this case.

In frame {\it b}) of Fig. 6, the entire region shown is allowed by
$B(b\to s\gamma )$, and in fact would be favored over the frame {\it a}) 
results by the CLEO central value, which
lies somewhat below the SM $B(b\to s\gamma )$ prediction. 
Meanwhile, much of the region shown in 
frame {\it c}) is excluded by $b\to s\gamma$, including all of the
cosmological MDM prefered region. Finally, in frame {\it d}),
the entire region is allowed by $B(b\to s\gamma )$. In fact, in this case,
the region around $m_{1/2}\sim 200$ GeV actually agrees with the central
value of the CLEO measurement of $B(b\to s\gamma )$, and overlaps
considerably with the cosmological MDM favored region!
The corresponding excluded
regions from the false vacuum constraint\cite{vacua} and 
$B(b\to s\gamma )$ for the $m_0\ vs.\ A_0$ plane for $m_{1/2}=200$ GeV
are shown in Fig. 7.

\subsection{Implications for Collider Experiments}

Next, we wish to draw some conclusions for future searches for mSUGRA at 
colliding beam experiments. Expectations for mSUGRA in the same 
$m_0\ vs.\ m_{1/2}$ planes have been worked out for the CERN LEP2
$e^+e^-$ collider\cite{bbmt},
various Fermilab Tevatron $p\bar p$ collider options\cite{tev}, 
and for the CERN Large Hadron 
Collider (LHC)\cite{bcpt}, a $pp$ collider expected to operate at 14 TeV.

We return attention to Fig. 6{\it a}. If the mSUGRA model represents nature
with parameters as in Fig. 6{\it a}, then the $B(b\to s\gamma )$ 
exclusion region wipes out most of the parameter space accessible to LEP2
SUSY searches. There are two exceptions\cite{bbmt}. 
There is a small region with
$m_0\agt 200$ GeV and $m_{1/2}\simeq 100$ GeV where charginos could be 
accessible to LEP2 operating around $\sqrt{s}\sim 190$ GeV. In this region,
the lightest neutralino $\tz_1$ can still be a good CDM candidate, since
$m_{\tz_1}\sim {M_Z\over 2}$-- thus, relic neutralinos can annihilate away via a $Z$
boson pole in the $s$-channel. The other possibility for LEP2 is to discover 
a light Higgs boson $h$ in the region beyond the $B(b\to s\gamma )$
exclusion contour. Much of the discovery reach of the Fermilab Main Injector 
($\int{\cal L}dt=2$ fb$^{-1}$) upgrade is also wiped out by the $b\to s\gamma$
constraint for the region $m_0\alt 180$ GeV. As with LEP2, there is some remaining region of accessibility
for seeing clean trileptons from $\tw_1\tz_2\to 3\ell$ around 
$m_0\agt 200$ GeV and $m_{1/2}\sim 120$ GeV. The TeV33 upgrade would 
be able to see in addition a large slice of parameter space beyond the
$b\to s\gamma$ excluded region up to $m_{1/2}\sim 275$ GeV for 
$m_0\alt 100$ GeV,
again via clean trileptons\cite{tev}. 
LHC would of course be able to scan well beyond the entire plane shown,
up to values of $m_{1/2}\sim 700-1000$ GeV with just 10 fb$^{-1}$ of data.
Since $B(b\to s\gamma )$ decreases with increasing $m_{1/2}$ and $m_0$,
however, the CLEO data prefer the region of parameter space accessible to LHC
experiments, rather than the regions accessible to LEP2 and Tevatron, in 
contrast to preferences from fine-tuning\cite{diego}.

Fig. 6{\it b} is entirely unconstrained by 
$B(b\to s\gamma )$. However, we do note that in this case the 
$B(b\to s\gamma )$ rate decreases with decreasing $m_{1/2}$ in the low
$m_{1/2}$ region. Hence, the CLEO data actually {\it prefer} the regions
accessible to LEP2 and Tevatron experiments, as do fine-tuning 
calculations.

For Fig. 6{\it c}, most of the plane shown is excluded by $B(b\to s\gamma )$,
so if nature chose these parameters, then LEP2 and Tevatron upgrades 
would see nothing, and the 
discovery of SUSY would have to wait for LHC, which could access the very heavy 
sparticle spectra that lie at parameter space points beyond 
$m_{1/2}\sim 350$ GeV. 

The parameter space of frame 6{\it d} is entirely allowed by 
$B(b\to s\gamma )$, and in fact the CLEO central value actually
agrees with the parameter space region around $m_{1/2}\sim 200$ GeV.
Since $B(b\to s\gamma )$ is decreasing below the SM
value with decreasing $m_{1/2}$, the low $m_{1/2}$ region accessible to
LEP2 and Tevatron upgrades is dis-favored by data. The region around
$m_0\sim 125$ and $m_{1/2}\sim 200$ is favored by $B(b\to s\gamma )$ and
by cosmology, and has no non-standard minima in the scalar potential.
In this favored region, unfortunately, neither SUSY nor Higgs particles 
would be accessible to LEP2 experiments. Only a fraction of this region
would be accessible to Tevatron upgrades via trilepton searches. However, 
the CERN LHC would enjoy huge supersymmetric signal rates in this region,
and in addition, direct production of sleptons would be visible as well.

\section{Conclusions}

We have performed a calculation of $B(b\to s\gamma )$ as a function
of the parameter space of the mSUGRA model. In doing so, we have included
several improvements over the usual leading-log treatment. We have included
corrections to the Wilson coefficients due to multiple scales in both the
SUSY and SM loop contributions. We have also included the published 
NLO terms in the ADM elements needed for running the Wilson coefficients
from a scale $Q=M_W$ down to $Q\sim m_b$. We have not included the
correction to $\gamma_{27}$, so that our final result is not NLL.
Preliminary results from the three loop calculation 
of $\gamma_{27}^{(1)}$ indicate it is only a small effect.
Finally, we have included the virtual and bremsstrahlung graphs in the 
evaluation of the $<s\gamma |O_i|b>$ matrix elements. The combination 
of all the above elements leads to a $B(b\to s\gamma )$ 
calculation with reduced uncertainty due to choice of scale.

We plot our results as a function of mSUGRA model parameter space, and
compare against recent results from the CLEO experiment. The comparison
leads to allowed and excluded regions of mSUGRA parameter space.
In particular, we note that for some mSUGRA parameter choices, the
mSUGRA $B(b\to s\gamma )$ prediction agrees better than the SM one.
The resulting constraints on parameter space are very strong, and indicate
that large regions for $\mu <0$ and for large $\tan\beta$ are excluded.
We compare briefly with expectations for collider experiments at LEP2,
Fermilab Tevatron, and CERN LHC. The $B(b\to s\gamma )$ constraint
rules out much of parameter space that would have been accessible to LEP2
and Tevatron experiments. However, we note that the region of 
parameter space most favored by $B(b\to s\gamma )$, cosmology
and standard minima of the scalar potential, around
$m_0\sim 125$ GeV, $m_{1/2}\sim 200$ GeV, $\tan\beta \sim 10$ and $\mu >0$
might be accessible to Fermilab MI or TeV33 searches for clean trileptons; if
not, the discovery of SUSY would have to wait until LHC experiments are
performed.


\acknowledgments

We thank J. Hewett and X. Tata for discussions. In addition, we thank
Manuel Drees for calculational comparisons which led to discovery of a bug
in the program used to generate results for an earlier version of this 
manuscript.
This research was supported in part by the U.~S. Department of Energy
under grant number DE-FG-05-87ER40319.

%

%
\newpage
%
%
\begin{figure}
\caption[]{
We plot the value of the Wilson coefficient $C_7(M_W)$ versus $m_{1/2}$,
where $m_0=m_{1/2}$, $\tan\beta =2$, $A_0=0$. In {\it a}), we take $\mu <0$
and in {\it b}) we take $\mu >0$. The various loop 
contributions to $C_7(M_W)$ are denoted on the plot. The label $\tilde q$
refers to the sum over $\tilde{q}=\tilde{u}_L,\tilde{u}_R,\tilde{c}_L,
\tilde{c}_R$ contributions.
}
\end{figure}
\begin{figure}
\caption[]{
Evolution of Wilson coefficients $C_i(Q )$ for $i=1-8$ from $Q =M_W$
to $Q =1$ GeV, for ($m_0,m_{1/2},A_0,\tan\beta ,sign(\mu )$=
$(100,100,0,10,-1$), where masses are in GeV units. Frame {\it a})
shows evolution to LO, while frame {\it b}) shows NLO evolution
(except $\gamma_{27}^{(1)}=0$).
}
\end{figure}
\begin{figure}
\caption[]{
A plot of the SM branching ratio $B(b\to s\gamma )$ versus scale choice
$Q$, where ${m_b\over 2}<Q <2 m_b$. We plot the LL result, and in addition,
our QCD-improved result. The theoretical error diminishes from $\sim 25\%$ to
$\sim 9\%$. We also plot the CLEO measured central value, as well as 
$1\sigma$ and 95\% CL limits on the experimental result.
}
\end{figure}
\begin{figure}
\caption[]{
Plot of contours of constant branching ratio $B(b\to s\gamma )$ in 
the $m_0\ vs.\ m_{1/2}$
plane, where $A_0=0$ and $m_t=170$ GeV. Each contour should be multiplied
by $10^{-4}$. Frame {\it a}) is for
$\tan\beta =2,\ \mu <0$, {\it b}) is for $\tan\beta =2,\ \mu >0$,
{\it c}) is for $\tan\beta =10,\ \mu <0$ and {\it d}) is for
$\tan\beta =10,\ \mu >0$.  
The regions labelled by TH (EX) are excluded by theoretical (experimental)
considerations. The dashed contour corresponds to the latest LEP2 limit
of $m_{\tw_1}>79$ GeV for a gaugino-like chargino.
}
\end{figure}
\begin{figure}
\caption[]{
Plot of contours of constant branching ratio $B(b\to s\gamma )$ in 
the $m_0\ vs.\ A_0$
plane, where $m_{1/2}=200$ GeV and $m_t=170$ GeV. 
Each contour should be multiplied
by $10^{-4}$. Frame {\it a}) is for
$\tan\beta =2,\ \mu <0$, {\it b}) is for $\tan\beta =2,\ \mu >0$,
{\it c}) is for $\tan\beta =10,\ \mu <0$ and {\it d}) is for
$\tan\beta =10,\ \mu >0$.  
The regions labelled by TH (EX) are excluded by theoretical (experimental)
considerations. 
}
\end{figure}
\begin{figure}
\caption[]{
Plot of contours of various constraints on the mSUGRA model in 
the $m_0\ vs.\ m_{1/2}$
plane, where $A_0=0$ and $m_t=170$ GeV. The frames are as in Fig. 4.
To the left of the contour marked by open circles is the region 
where minima occur in the mSUGRA scalar potential that are deeper than the
standard one.
The region below the dashed contour is excluded by the LEP2 limit
that $m_{\tw_1}>79$ GeV.
The region to the right of the solid contour labelled $\Omega h^2$ is 
excluded because the universe would be younger than 10 billion years old
($\Omega h^2 >1$). The region between the dot-dashed contours is favored
by the cosmological mixed dark matter scenario, where $0.15<\Omega h^2<0.4$.
Finally, the region to the lower-left of the solid contour labelled
$b\to s\gamma$ in frame {\it a}) is excluded at 95\% CL by the 
analysis of this paper. The entire region in frame {\it b}) is allowed
by $b\to s\gamma$, while almost the entire region in frame {\it c}) is excluded
by $b\to s\gamma$. Finally, the entire region shown in 
frame {\it d}) is again allowed by the $b\to s\gamma$ constraint.
}
\end{figure}
\begin{figure}
\caption[]{
Plot of contours of various constraints on the mSUGRA model in 
the $m_0\ vs.\ A_0$
plane, where $m_{1/2}=200$ GeV and $m_t=170$ GeV. The frames are as in Fig. 5.
To the left of the contour marked by open circles is the region 
where minima occur in the mSUGRA scalar potential that are deeper than the
standard one.
The entire regions in frames {\it a}), {\it b}) and {\it d}) are allowed
by $b\to s\gamma$, while the entire region in frame {\it c}) is excluded
by $b\to s\gamma$. 
}
\end{figure}

\vfill\eject


\centerline{\epsfbox{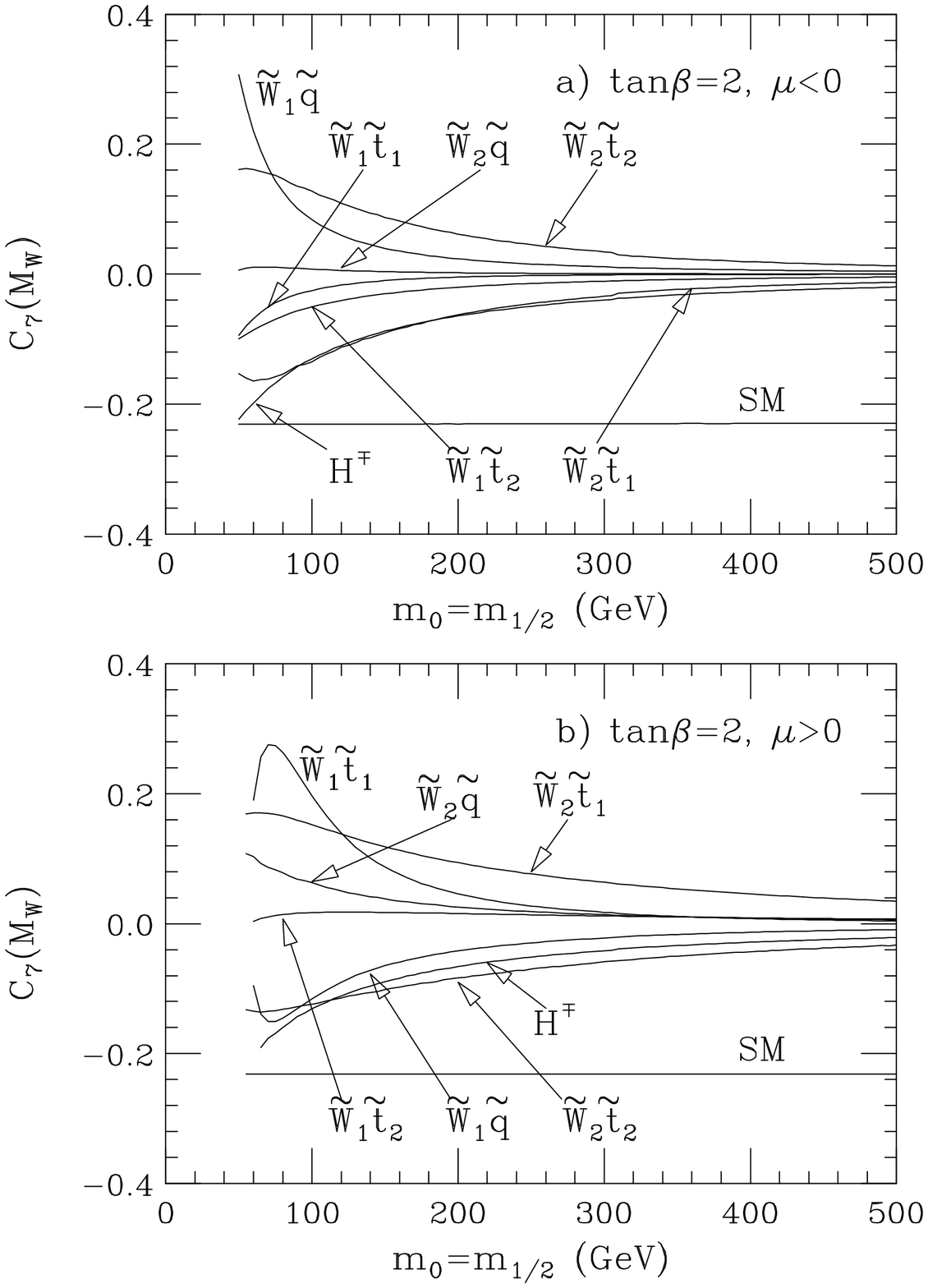}}
\bigskip
\centerline{Fig.~1}
\vfill\eject

\centerline{\epsfbox{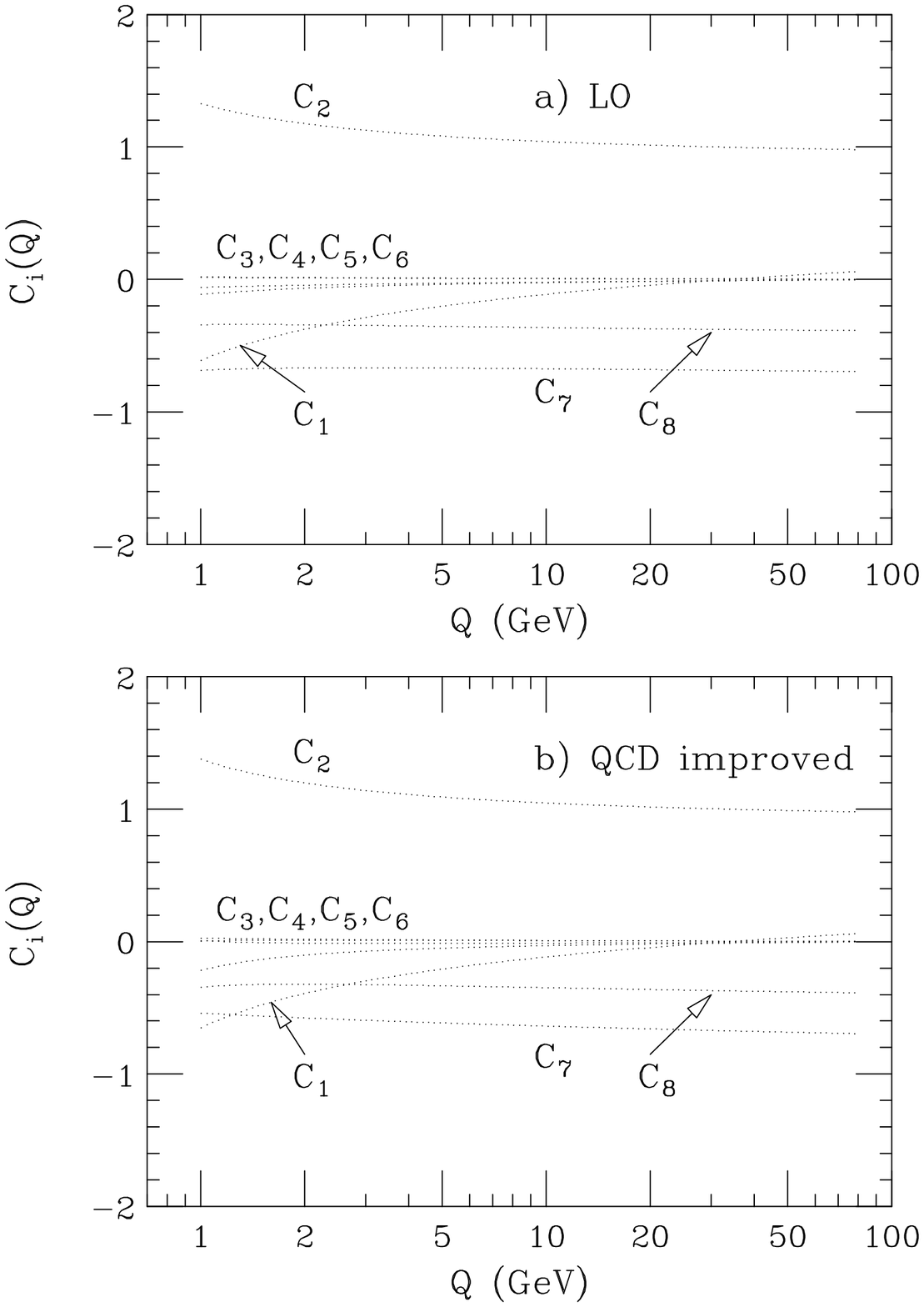}}
\bigskip\bigskip
\centerline{Fig.~2}
\vfill\eject

\centerline{\epsfbox{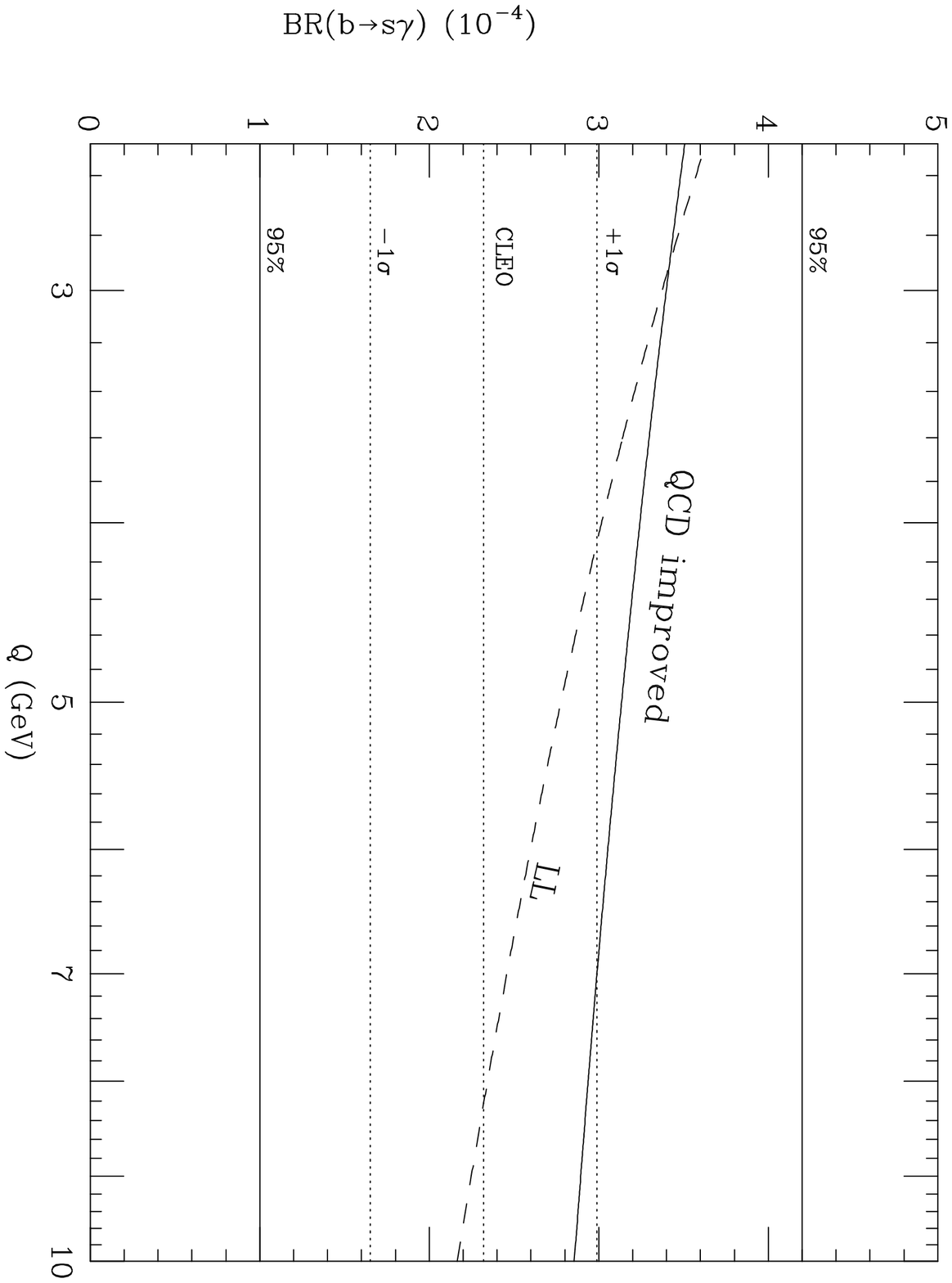}}
\bigskip\bigskip
\centerline{Fig.~3}
\vfill\eject

\centerline{\epsfbox{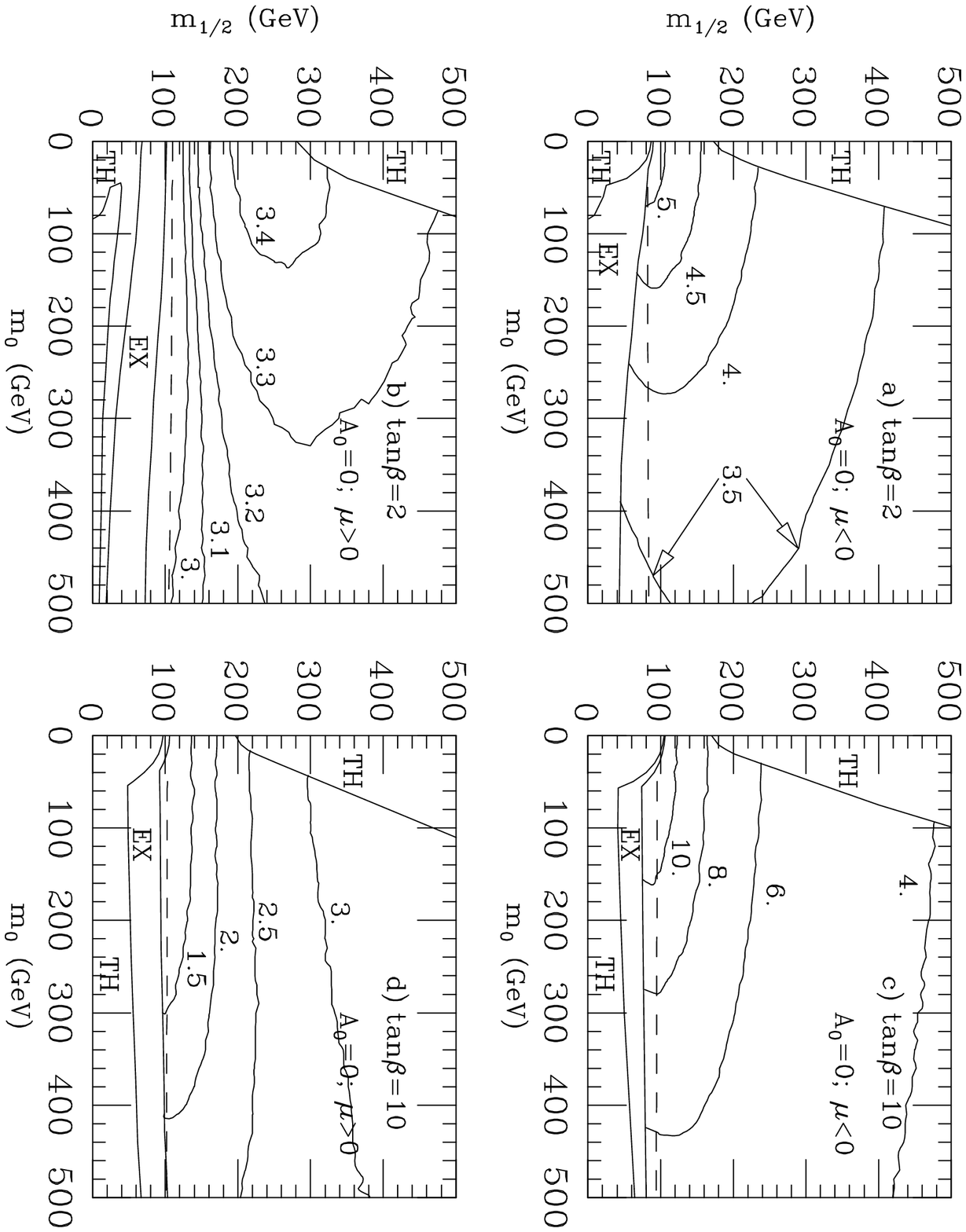}}
\bigskip\bigskip
\centerline{Fig.~4}
\vfill\eject

\centerline{\epsfbox{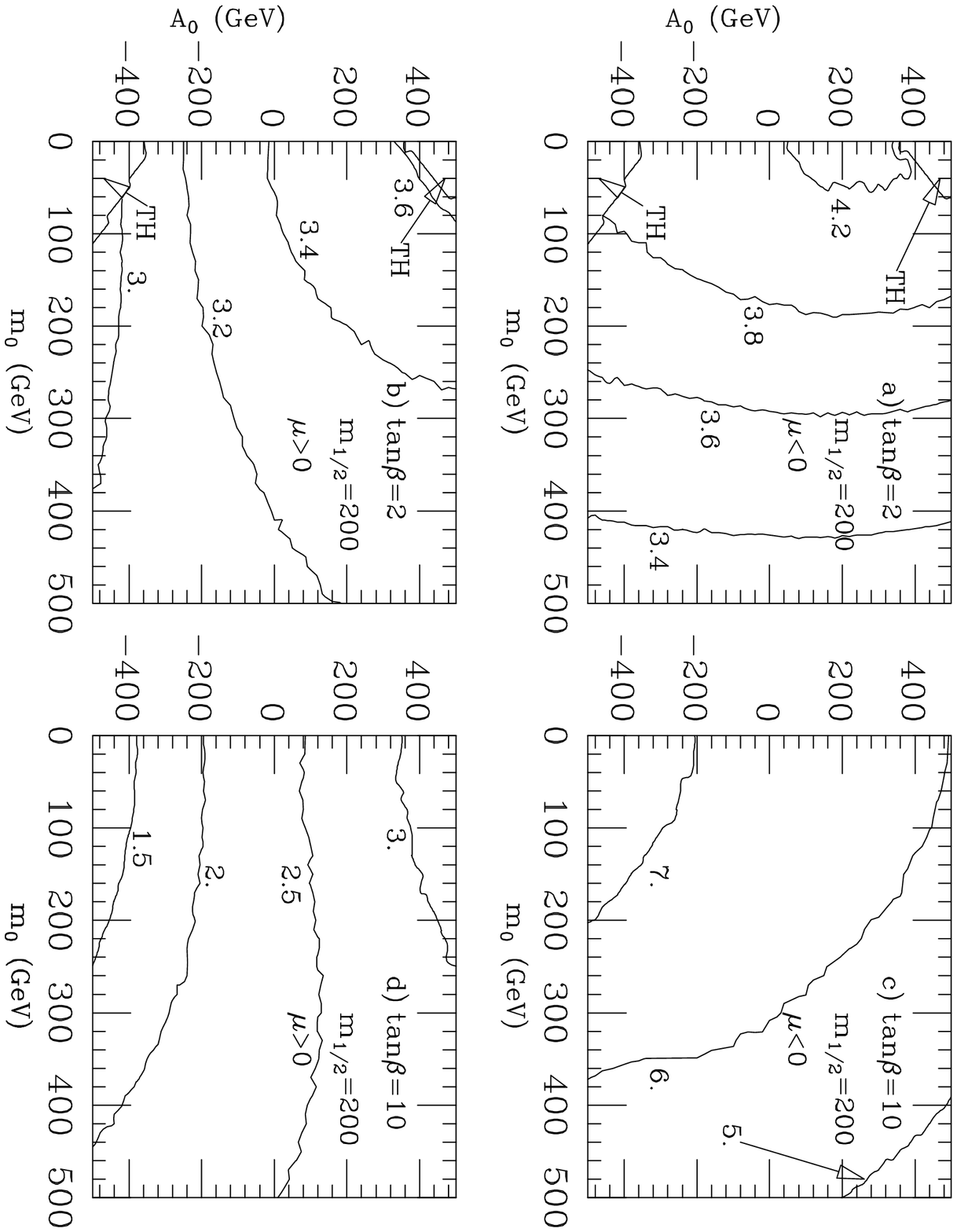}}
\bigskip\bigskip
\centerline{Fig.~5}
\vfill\eject

\centerline{\epsfbox{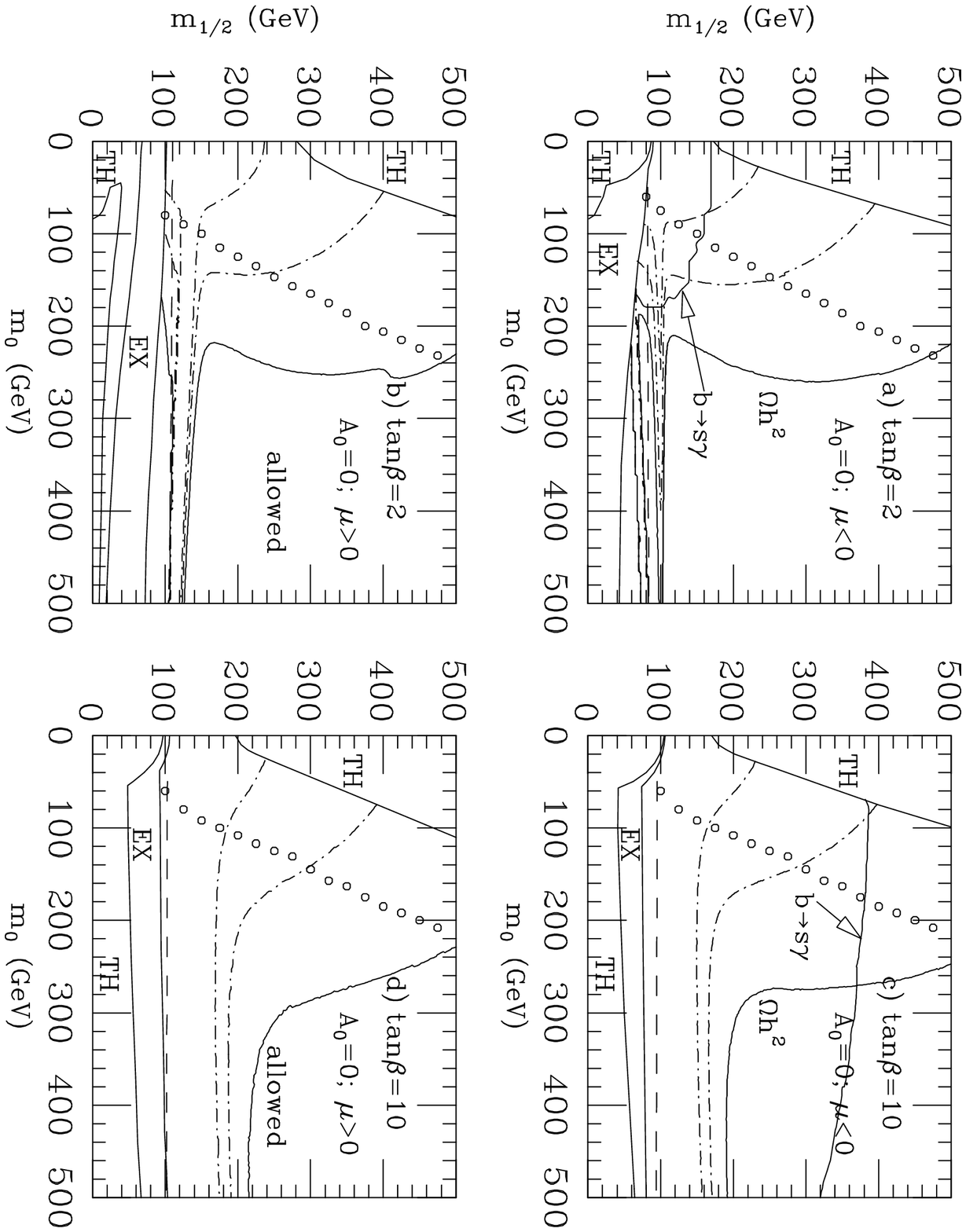}}
\bigskip\bigskip
\centerline{Fig.~6}
\vfill\eject

\centerline{\epsfbox{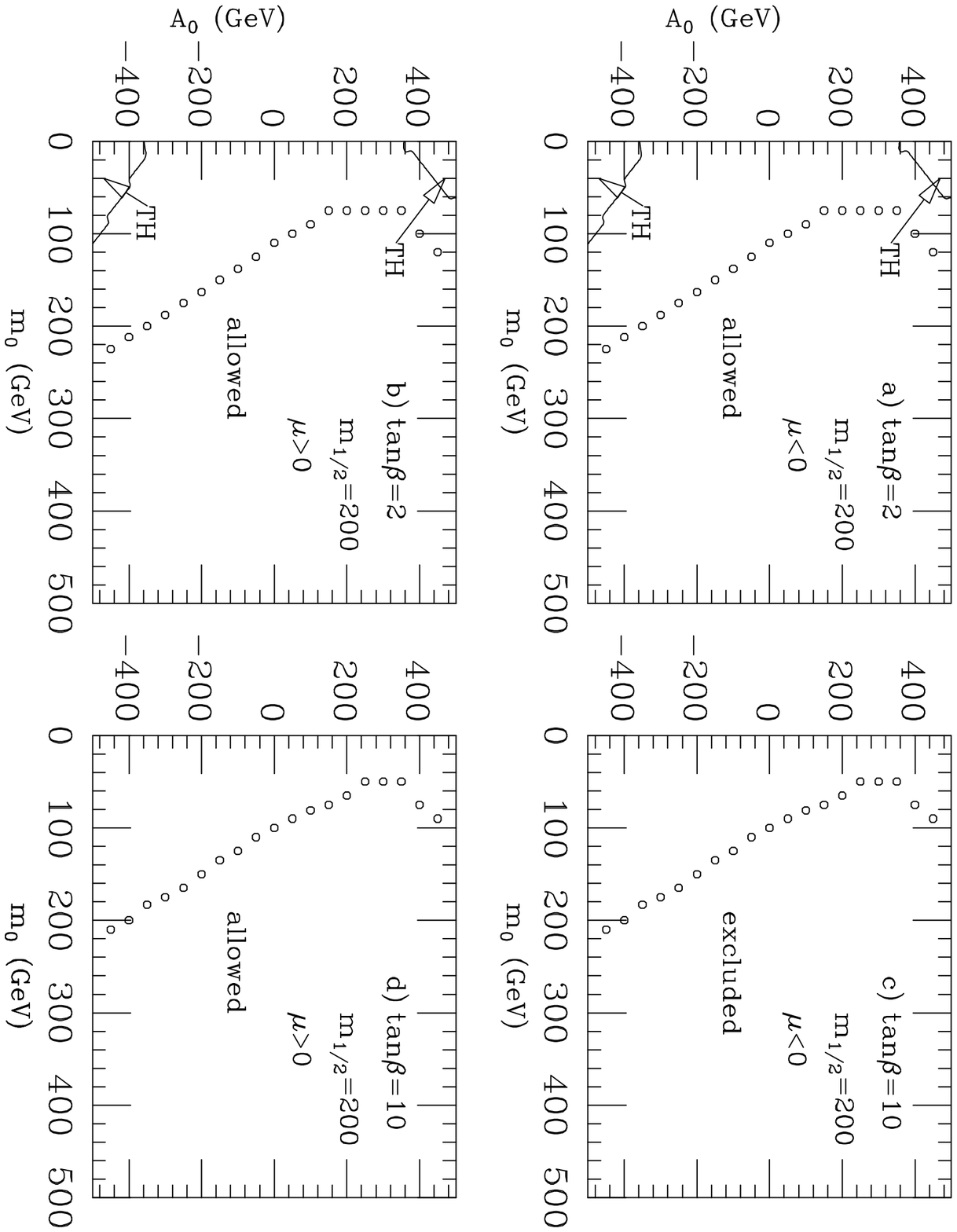}}
\bigskip\bigskip
\centerline{Fig.~7}
\vfill\eject


\begin{references}
\bibitem{haber}  See, {\it e.g.} 
H. Haber in {\it Woodlands Superworld}, hep-ph/9308209 (1993). 
%
\bibitem{drees} See {\it e.g.} M. Drees and S. Martin in 
{\it Electroweak Symmetry Breaking and New Physics at the TeV Scale}, 
edited by T. Barklow, S. Dawson, H~. Haber and J.~Seigrist, (World Scientific)
1995; see also J. Amundson {\it et. al.}, hep-ph/9609374 (1996).
%
\bibitem{sugra} A. Chamseddine, R. Arnowitt and P. Nath, 
Phys. Rev. Lett. {\bf 49}, 970 (1982);
R. Barbieri, S. Ferrara and C. Savoy, Phys. Lett. {\bf B119}, 343 (1982);
L.J. Hall, J. Lykken and S. Weinberg, Phys. Rev. {\bf D27}, 2359 (1983).
%
\bibitem{lep2} R. Cavanaugh (ALEPH Collaboration), 
seminar at Florida State University on {\it New Results from LEP2},
September, 1996.
%
\bibitem{cosmo} For a recent review, see
G. Jungman, M. Kamionkowski and K. Griest, 
Phys. Rep. {\bf 267}, 195 (1996). See also M. Drees and M. Nojiri,
Phys. Rev. {\bf D47}, 376 (1993). The result shown here are from
H. Baer and M. Brhlik, Phys. Rev. {\bf D53}, 597 (1996). Further 
references are included in these reports.
%
\bibitem{vacua} J. A. Casas, A. Lleyda and C. M\~unoz, 
Nucl. Phys. B{\bf 471}, 3 (1996);
H. Baer, M. Brhlik and D. Casta\~no, Phys. Rev. D. (in press), 
hep-ph/9607465 (1996).
%
\bibitem{hewett}  See talk by J. Hewett presented at SLAC Summer
Inst. on Particle Physics, SLAC-PUB-6521 (1993), hep-ph/9406302.
%
\bibitem{cleo} M. S. Alam {\it et. al.}, (CLEO Collaboration),
Phys. Rev. Lett. {\bf 74}, 2885 (1995).
%
\bibitem{higdoub} B. Grinstein, R. Springer and M. Wise, 
Nucl. Phys. B{\bf 339}, 269 (1989); 
J. Hewett, Phys. Rev. Lett. {\bf 70}, 1045 (1993); 
V. Barger, M. Berger and R. Phillips, Phys. Rev. Lett. {\bf 70}, 1368 (1993); 
M. Diaz, Phys. Lett. B{\bf 304}, 278 (1993);
N. Deshpande, K. Panose and J. Trampetic, Phys. Lett. B{\bf 308}, 322 (1993).
%
\bibitem{masiero} S. Bertolini, F. Borzumati, A. Masiero and G. Ridolfi,
Nucl. Phys. B{\bf 353}, 591 (1991).
%
\bibitem{susybsg} R. Barbieri and G. F. Giudice, 
Phys. Lett. B{\bf 309}, 86 (1993); J. Lopez, D. Nanopoulos and G. Park, 
Phys. Rev. D{\bf 48}, 974 (1993); N. Oshimo, Nucl. Phys. B{\bf 404}, 20 (1993);
R. Garisto and J. Ng, Phys. Lett. B{\bf 315}, 372 (1993); 
M. Diaz, Phys. Lett. B{\bf 322}, 207 (1994);
Y. Okada, Phys. Lett. B{\bf 315}, 119 (1993);
F. Borzumati, Zeit. f\"ur Physik C{\bf 63}, 291 (1994);
P. Nath and R. Arnowitt, Phys. Lett. B{\bf 336}, 395 (1994);
G. Kane, C. Kolda, L. Roszkowski and J. Wells, 
Phys. Rev. D{\bf 49}, 6173 (1994);
F. Borzumati, M. Drees and M. Nojiri, Phys. Rev. D{\bf 51}, 341 (1995);
V. Barger, M. Berger, P. Ohmann and R. Phillips, 
Phys. Rev. D{\bf 51}, 2438 (1995);
F. Bertolini and F. Vissani, Zeit. f\"ur Physik C{\bf 67}, 513 (1995);
J. Lopez, D. Nanopoulos, X.~Wang and A.~Zichichi, 
Phys. Rev. D{\bf 51}, 147 (1995);
J. Wu, R. Arnowitt and P. Nath, Phys. Rev. D{\bf 51}, 
1371 (1995); 
B. de Carlos and J. A. Casas, Phys. Lett. B{\bf 349}, 300 (1995) 
and ERRATUM-{\it ibid} B{\bf 351}, 604 (1995).
%
\bibitem{gh} C. Greub and T. Hurth, SLAC-PUB-7267 (1996), hep-ph/9608449.
%
\bibitem{llqcd} M. A. Shifman, A. I. Vainshtein and V. I. Zacharov, 
Phys. Rev. D{\bf 18}, 2583 (1978);
S. Bertolini, F. Borzumati and A. Masiero, 
Phys. Rev. Lett. {\bf 59}, 180 (1987);
N. G. Deshpande, G. Eilam, P. Lo and J. Trampetic, 
Phys. Rev. Lett. {\bf 59}, 183 (1987);
B. Grinstein, R. Springer and M. Wise, Phys. Lett. B{\bf 202}, 138 (1988);
R. Grigjanis, P. J. O'Donnel and M. Sutherland, 
Phys. Lett. B{\bf 213}, 355 (1988) and Phys. Lett. B{\bf 224}, 209 (1989);
G. Cella, G. Curci, G. Ricciardi and A. Vicere, 
Phys. Lett. B{\bf 248}, 181 (1990);
M. Misiak, Phys. Lett. B{\bf 269}, 161 (1991) and 
Nucl. Phys. B{\bf 393}, 23 (1993);
M. Ciuchini, E. Franco, G. Martinelli, L. Reina and L. Silvestrini,
Phys. Lett. B{\bf 316}, 127 (1993);
M. Ciuchini, E. Franco, L. Reina and L. Silvestrini,
Nucl. Phys. B{\bf 421}, 41 (1993).
%
\bibitem{pokorski} A. J. Buras, M. Misiak, M. M\"unz and S. Pokorski, 
Nucl. Phys. B{\bf 424}, 374 (1994).
%
\bibitem{cg} P. Cho and B. Grinstein, Nucl. Phys. B{\bf 365}, 279 (1991).
%
\bibitem{anlauf} H. Anlauf, Nucl. Phys. B{\bf 430}, 245 (1994).
%
\bibitem{ciuchini} M. Ciuchini, E. Franco, G. Martinelli and  L. Reina,
Nucl. Phys. B{\bf 415}, 403 (1994).
%
\bibitem{misiak1} M. Misiak and M. M\"unz, Phys. Lett. B{\bf 344}, 308 (1995).
%
\bibitem{misiak2} K. G. Chetyrkin, M. Misiak and M. M\"unz, 
in preparation; talk given by M. Misiak at ICHEP96, Warsaw, Poland,
July, 1996.
%
\bibitem{soares} J. M. Soares, Phys. Rev. D{\bf 49}, 283 (1994).
%
\bibitem{ghw} C. Greub, T. Hurth and D. Wyler, 
Phys. Lett. B{\bf 380}, 385 (1996) and Phys. Rev. D{\bf 54}, 3350 (1996).
%
\bibitem{brem} A. Ali and C. Greub, Z. Phys. C{\bf 49}, 431 (1991),
Phys. Lett. B{\bf 259}, 182 (1991), {\bf 287}, 191 (1992) and 
{\bf 361}, 146 (1995);
Z. Phys. C{\bf 60}, 433 (1993); N. Pott, Phys. Rev. D{\bf 54}, 938 (1996);
%
\bibitem{bcmpt} H. Baer, C. H. Chen, R. Munroe, F. Paige and X. Tata,
Phys. Rev. D{\bf 51}, 1046 (1995).
%
\bibitem{desh} For discussion, see N.G. Deshpande, X. G. He and J. Trampetic,
Phys. Lett. {\bf B367}, 362 (1996).
%
\bibitem{diego} G. Anderson and D. Casta\~no, Phys. Rev. D{\bf 53}, 2403 (1995).
%
\bibitem{bbmt} H. Baer, M. Brhlik, R. Munroe and X. Tata,
Phys. Rev. D{\bf 52}, 5031 (1995).
%
\bibitem{tev} H. Baer, C. H. Chen, C. Kao and X. Tata,
Phys. Rev. D{\bf 52}, 1565 (1995); H. Baer, C. H. Chen, F. Paige and X. Tata,
Phys. Rev. D (in press).
%
\bibitem{bcpt} H. Baer, C. H. Chen, F. Paige and X. Tata,
Phys. Rev. D{\bf 52}, 2746 (1995) and Phys. Rev. D{\bf 53}, 6241 (1996).
%
%

\end{references}
\end{document}